\documentclass[12pt]{article}
\usepackage[english]{babel}
\usepackage{amsmath}
\usepackage{times}
\usepackage[body={6.5in,9.5in}]{geometry}

\usepackage{amsmath}
\usepackage{graphicx}
\usepackage{subfigure}
\usepackage{color}

\title{\it Covariant  Magnetic Connection Hypersurfaces}
\author{{\it F. Pegoraro}\\~~\\
{\small  Dipartimento di Fisica, Universit\`a di Pisa, Italy}\\{\small francesco.pegoraro@unipi.it}\\{\small ~~}\\{\small Accepted for publication in {\it Journal of Plasma Physics}}}
\date{}

\begin{document}
\maketitle

\begin{abstract}
In the single fluid, nonrelativistic, ideal-Magnetohydrodynamic (MHD)  plasma description magnetic field lines  play a fundamental role by defining  dynamically preserved ``magnetic connections'' between plasma elements. Here we show how the concept of magnetic connection  needs to  be generalized in the case of a relativistic MHD description where we require covariance under arbitrary Lorentz  transformations. This  is performed by defining 2-D {\it magnetic connection hypersurfaces} in the 4-D Minkowski space. This generalization  accounts for the loss of simultaneity between spatially separated events in different frames and is expected to provide a powerful insight into  the 4-D geometry of electromagnetic fields when ${\bf E} \cdot {\bf B} = 0$.    
\end{abstract}

{\it Pacs:} {\small 52.27.Ny,\, 
03.50.De, \, 52.35.Vd
}

\section{Introduction}\label{intr}

The dynamics of large scale relativistic plasma configurations plays an important role in our understanding of high energy astrophysical phenomena such as, just to mention a recently discovered one,   the flaring of the Crab nebula \cite{Tav}.
\,
Even without including general relativistic effects, as would  be the case,  e.g., in the neighbourhood of a black hole (see for example the system of equations investigated  by \cite{Koide}), the phenomena we need to describe involve velocities close to the speed of light  and internal energies that can be larger than the electron rest  mass energy.
Furthermore relativistic plasmas with very large energy densities  have been produced in the laboratory  in laser  plasma experiments and it has been stressed that  such experiments can help us to understand high energy astrophysical phenomena (see e.g., \cite{bul}).

With this in mind, several  concepts  that have been introduced for nonrelativistic plasmas need to  be extended to relativistic regimes.  In such  a generalization   space and time properties are necessarily  combined since the basic invariance properties of the matter equations are now given in terms  of the Lorentz  group of transformations between different  reference frames. This is particularly important since, in the presence of very large velocity differences between different parts of the plasma configuration, there may not be a clear way  of defining 
 a  preferred reference frame on physics grounds. In addition, the observer reference frame may move with a relativistic velocity with respect to the plasma under observation and thus observe as simultaneous events that are not simultaneous in the  plasma frame.
\medskip

 For phenomena occurring on macroscopic scales, i.e.  on space and time scales large on the characteristic  microscopic scales of the  particle dynamics,  the single fluid nonrelativistic MHD plasma description has been  extended (see  \cite{Li,Ani}) and used in numerical simulations (see e.g., \cite{Mign})  so as 
  to include relativistic fluid  velocities and relativistic  internal energy densities.  A Hamiltonian reformulation  of ideal relativistic MHD dynamics  in terms of noncanonical variables has been recently derived in \cite{D'A}.
 \, 
In this process of  generalization  a number of basic phenomena of nonrelativistic MHD, such as e.g. magnetic reconnection, have been reconsidered in  relativistic  plasma regimes both  in the laboratory (see  \cite{Askar,nilson}), and in  astrophysics (see \cite{HessZen,HoshZen}). 
In particular in the astrophysical context  relativistic magnetic reconnection has been considered mostly as a mechanism of energy conversion,  usually choosing a preferred frame of reference, possibly thought  of as an
``average  comoving frame'', i.e. as a frame in which the plasma region under consideration is globally  at rest.
As mentioned above, such an approach may not be fully unambiguous  in  situations where very large velocity  relativistic variations can be present between different plasma regions. This is so  in particular from the observational point of view 
when describing magnetic reconnection structures since magnetic fields and electric fields  are transformed one into the other when seen in a  Lorentz boosted reference frame. \\
Thus an important point in the relativistic extension of the MHD  plasma description is to provide a frame independent definition of magnetic reconnection. However such a definition  is neither  obvious  from a theoretical nor  from an observational point of view since, as already mentioned,   the distinction between electric and magnetic fields is frame dependent and the tracing of field lines, which are only defined in coordinate space at fixed  time, is also frame dependent due to the violation of simultaneity in different reference frames of events at different spatial locations. 
\medskip

Although a clearcut definition of magnetic reconnection is not simple  to formulate even for a non relativistic plasma, its common definition is not simply limited to the fact that magnetic energy  is  converted to kinetic  and/or internal plasma energy  but refers to the  local violation of the magnetic topology and, in particular,  to the {local}  breaking of the structure of magnetic connections. \\ Magnetic connections are defined by the  fundamental property of ideal MHD (see \cite{Newc}) that if two plasma elements, moving with plasma in a smooth flow,  are connected at time $t$ by a magnetic field line  then at any following time there exists a magnetic field line that connects them.  This property is the conceptual basis from  which the expressions that the magnetic field is frozen in the plasma and that field lines move with the plasma are derived.\\
Thus in order to define magnetic reconnection in a covariant way we must first obtain a covariant definition of magnetic connections. Again, such a definition is not {\it a priori} obvious because 
of two already mentioned related reasons:   the distinction between electric and magnetic fields and the very concept of field lines are frame dependent.  This point  was  explicitly addressed in \cite{FPEPL}  where   it was shown that the  covariant formulation of magnetic connections  can be restored by means of a {\it time resetting} projection along the trajectories of the plasma elements. This projection is consistent with the ideal Ohm's law and   compensates  for the loss of simultaneity in different reference frames between spatially separated events.  
\medskip

In the present paper we address this same issue again and  show  that the time resetting  along the trajectories of the fluid elements introduced in \cite{FPEPL} is essentially equivalent to a redefinition of the geometrical object that  we use in order to define magnetic connections. We argue that, while in 3-D (coordinate) space  magnetic connections are defined by  1-D curves (field lines), in the 4-D Minkowski space they are defined by 2-D hypersurfaces that are generated by a suitably defined  magnetic (space-like) 4-vector field and by the  velocity  (time-like)  4-vector field of the plasma.
\\
In fact, following a somewhat different line of approach from the one  adopted in \cite{FPEPL}, we show that, if  the electromagnetic (e.m) field tensor satisfies an ideal Ohm's law,  it exhibits special geometrical  properties that are simply the consequence of the homogeneous Maxwell's equations  and that make it possible  to define such 2-D hypersurfaces so  that, if in a given frame two plasma elements  in 4-D Minkowski space lie on the same 2-D hypersurface, they do so in any  other reference frame. 
\\ We call these  2-D hypersurfaces   (with one space-like and one time-like tangent vector field) {\it Covariant  Magnetic Connection Hypersurfaces}, or connection hypersurfaces for short.  The standard  magnetic connections in 3-D space can then be recovered in any chosen reference frame by taking sections of these surfaces at a fixed (in that frame) time.  We stress that  these 2-D hypersurfaces bear no relation to the 3-D magnetic surfaces  of nonrelativistic MHD that, if generalized to 4-D Minkowski space, would involve 3-D ``volumes''.

The present paper stops at this result, just   after noting that the violation of the ideal Ohm's law leads to a violation of the geometrical properties of the e.m. field tensor that make it possible to define the connection hypersurfaces. Thus in this  4-D framework magnetic reconnection, caused  by a local violation of the ideal Ohm's law, can be interpreted in a frame independent way as a local  ``piercing and merging'' of connection hypersurfaces. These  lose  their  identity only locally, in exactly the same way as  magnetic field  lines do in the standard 3-D space  setting.  The  physical and observational consequences of this definition will be  investigated in detail in a later paper.   \\ However, even remaining within the validity of the ideal Ohm's law, i.e.  without allowing for magnetic reconnection to occur, important open questions remain to be investigated: in particular how to generalize the study of the topological properties such as, e.g., field line braiding (see e.g., \cite{berger}), that have been investigated within a fixed frame 3-D description to the study  of the properties of connection hypersurfaces in 4-D Minkowski space. 
In the present paper only some very general properties of the  magnetic helicity 4-vector field are discussed and  are shown to allow us to define a Lorentz-scalar  Lagrangian invariant that is  advected by the plasma motion.

Before entering the detailed derivation  of the covariant  connection hypersurfaces  we stress  that  their definition only requires that an ideal Ohm's law be valid, supplemented  by  the homogeneous Maxwell's equations.
The inhomogeneous Maxwell's equations, that relate  the e.m. field tensor  to the charge and current densities  and that thus determine the field dynamics from the plasma dynamics,
 are not directly involved in  the definition of the connection hypersurfaces which,    in this sense, are  more general than relativistic MHD and thus apply under more general conditions.. Depending on the plasma description adopted,  the connection hypersurfaces can either relate to the single fluid description or to a selected species in the plasma, generally the lighter one.
Physically,  the main assumption  that is made is that kinetic effects can be neglected in the chosen regime  for this lighter species and that a fluid velocity can be defined, independently of whether 
it is a single fluid velocity, as in MHD, or  e.g.,  the electron velocity.  We also note  (see e.g.,  \cite{FPPoP})  that electron inertia effects and electron thermal effects (for an isotropic and isentropic thermodynamic
closure) can be included by a suitable redefinition of the electromagnetic field tensor. In fact, this redefined field tensor   obeys  an ideal Ohm's law and  a set of equations analogous in form to the homogeneous Maxwell equations. 
 On the other hand  dissipative  effects, either resistive  or arising from the ``friction term'' due to incoherent high frequency radiation  in   fully relativistic regimes,  can lead to violation of the ideal Ohm's law, 
in particular in the presence  of a nonlinear plasma dynamics that  leads  to the formation  of smaller and smaller space and time scales. 
If these effects are local, they provide the  local  breaking and merging of the connection hypersurfaces involved in magnetic reconnection.
Finally  the case of an electron-positron plasma where there are two light species,  and that is quite important for astrophysics,   would require within the present framework  the additional assumption that  both  species  satisfy   an ideal OhmÕs law (not necessarily the same).

\section{Ideal Ohm's law}\label{ROL}

An important feature of the ideal 3-D Ohm's law 
\begin{equation} \label{int1}  {\bf E} + {\bf v}\times {\bf B}/c =0, \quad \Rightarrow {\bf E}\cdot {\bf B} =0, \end{equation}  with ${\bf v}$  the 3-D plasma fluid velocity field  and ${\bf E}$ and  ${\bf B}$ the electric and the magnetic fields, is that it is in no sense restricted to a nonrelativistic plasma regime
or to a preferred reference frame.  In fact it can be written (unmodified) in the fully covariant form (see e.g., \cite{Ged})
\begin{equation}\label{int2}
{\bf F}_{\mu\nu}{\bf u}^\nu = 0,
\end{equation}
where ${\bf F}_{\mu\nu}$ is the e.m.  field tensor, ${\bf u}^\mu$ is a timelike 4-vector which we interpret as the fluid velocity  4-vector field of the plasma  (or of the plasma species with respect to which the magnetic field is frozen, see e.g. the generalized formulation given in  \cite{FPPoP,AsCo1,AsCo2}).\\
From Eq.(\ref{int1})  and Faraday's equation $\nabla\times {\bf E} \, + \, (1/c)\, \partial{\bf B} /\partial t  = 0$,  the 3-D magnetic equation 
\begin{equation}\label{int3}
\partial{\bf B} /\partial t  - \nabla\times ({\bf v} \times {\bf B}) =0
\end{equation}
 follows,  together with  the 3-D connection theorem \cite{Newc} mentioned in the Introduction:
 if at $t=0$  we have $d {\bf l}\times{\bf B}=0$, where $d {\bf l}$  is the the vector field tangent  to a curve connecting two plasma elements, i.e., if the two elements are connected by a magnetic field line, then  
$\,d {\bf l}\times{\bf B}=0\,$ for all $t$ since 
\begin{equation}\label{int4} 
\frac{d}{dt}{\left(d
 {\bf l}\times{\bf B}\right)}= -{\left(d
{\bf l}\times{\bf B}\right)}\left({\nabla}\cdot{\bf v}\right)
-\left[{\left(d{ \bf l}\times{\bf B}\right)}\times{\nabla}\right]{\bf v}. \end{equation} 
Here $d/dt$ is the Lagrangian time derivative along the plasma element motion.\\
While the ideal Ohm's law (\ref{int1}) can be set in an explicitly covariant  form (\ref{int2}), the   interpretation of Eq.(\ref{int4}) in terms of the conservation of magnetic connections cannot be  directly transferred to a different reference frame, as can be seen  from the fact that  a Lorentz boost  will in general add a time component to the transformed   vector field $d {\bf l}'$ so that it will no longer be possible to interpret it as  the vector field   tangent to a curve in 3-D (coordinate) space. \\ However the simple fact that Ohm's law is fully covariant suggests  that it must be possible to reformulate the connection theorem in a frame independent way.

\subsection{Lichnerowicz-Anile  representation}\label{LAR}

In contrast  to \cite{FPEPL}, here we adopt the   two 4-vector fields representation  \cite{Li,Ani,D'A}  of the e.m. field tensor  ${\bf F}_{\mu\nu}$
\begin{equation}\label{2} {\bf F}_{\mu\nu}  =  {\varepsilon_{\mu\nu\lambda\sigma} {\bf b}^{\lambda}{\bf u}^{\sigma}}\,  + \,  {[{\bf u}_{\mu}{\bf e}_{\nu} - {\bf u}_{\nu}{\bf e}_{\mu}]} \, , \end{equation}
where  ${\bf b}^\mu$  is the {\it 4-vector magnetic  field} and ${\bf e}_\mu$  is the {\it 4-vector electric field}, 
with  ${\bf u}^\mu {\bf e}_\mu =0$ and ${\bf u}_\mu {\bf b}^\mu =0$. The  4-vectors ${\bf e}_\mu$ and ${\bf b}^\mu$ are related to the standard electric and magnetic fields ${\bf E}$ and ${\bf B}$ in 3-D space by
\begin{equation}\label{2-b}
{\bf b}^\mu = \gamma({\bf B}  + {\bf E} \times {\bf v} \,  , \, {\bf B} \cdot  {\bf v}\  ), 
\end{equation}
and 
\begin{equation}\label{2-c}
{\bf e}_\mu = \gamma({\bf E} + {\bf v} \times {\bf B} \,  , \, -{\bf E} \cdot  {\bf v}   ) ,
\end{equation}
with ${\bf e}_\mu {\bf b}^\mu =  {\bf E}  \cdot {\bf B}$. \, We have adopted  the Minkowski metric tensor  \, $\eta_{\mu\nu}$ defined by $(+,+,+,-)$ and  normalized 3-D velocities ${\bf v}$ to the speed of light:  $\gamma$ is the relativistic Lorentz factor and  $ {\bf u}^\mu = \gamma ( {\bf v},1)$ and $~{\bf u}_\mu {\bf u}^\mu = -1$. The orthogonality conditions  ${\bf u}^\mu {\bf e}_\mu = {\bf u}_\mu {\bf b}^\mu =0$  make this representation  unique.\,
In the following we will call this representation  the Lichnerowicz-Anile (LA)  representation.
The LA representation
 is physically convenient as it allows us to separate  covariantly   the magnetic and the  electric  parts of the e.m. field tensor  on dynamical grounds, i.e. relative to  the   plasma   velocity 4-vector field ${\bf u}^\mu$.  In the  local rest frame of a  plasma element   the time components of ${\bf e}_\mu$ and of ${\bf b}^\mu$ vanish, while their space components reduce to the standard 3-D electric and magnetic fields. 
\\
A corresponding representation holds for the dual tensor ${\bf G }^{\mu\nu} \equiv  \varepsilon^{\mu\nu\alpha\beta} {\bf F}_{\alpha\beta}/2$ with $ {\bf e}_\mu $ and $ {\bf b}^\mu $ interchanged. \, Thus:
\begin{equation}  {\bf G }^{\mu\nu}  =  {\varepsilon^{\mu\nu\lambda\sigma} {\bf u}_{\lambda}{\bf e}_{\sigma}}\,  + \,  {[{\bf u}^{\mu}{\bf b}^{\nu} - {\bf u}^{\nu}{\bf b}^{\mu}]},  \quad  {\rm with} \,
{\bf e}_\mu = {\bf F}_{\mu\nu} {\bf u}^\nu  \quad {\rm and} \quad  {\bf b}^\mu = {\bf G }^{\mu\nu}  {\bf u}_\nu.  \label{defin}\end{equation}

If the ideal Ohm's law ${\bf F}_{\mu\nu}{\bf u}^\nu =0$   holds, the electric 4-vector ${\bf e}_\mu$ vanishes, the tensors ${\bf F}_{\mu\nu}$ and ${\bf G }^{\mu\nu}$ have rank two  and can be written as 
\begin{equation}\label{2b} {\bf F}_{\mu\nu}  =  {\varepsilon_{\mu\nu\lambda\sigma} {\bf b}^{\lambda}{\bf u}^{\sigma}}\, , \qquad  \,  {\bf G }^{\mu\nu}  = {[{\bf u}^{\mu}{\bf b}^{\nu} - {\bf u}^{\nu}{\bf b}^{\mu}]} \, , \end{equation}
with 
\begin{equation}\label{2c} {\bf F}_{\mu\nu} {\bf b}^\nu =  {\bf F}_{\mu\nu} {\bf u}^\nu  =0,  \ \end{equation} 
\begin{equation}\label{2d}  {\bf F}_{\mu\nu} {\bf G }^{\nu\mu} = 0 \,  \rightarrow   {\bf E}  \cdot {\bf B} =0, \qquad {\rm and} \quad  {\bf b}_\mu {\bf b}^\mu  =  {\bf G }_{\mu\nu}  {\bf G }^{\nu\mu}/2  = {\bf F}_{\mu\nu}  {\bf F}^{\nu\mu}/2  . \end{equation}
In this case we can use ${\bf e}_\mu =0$ in order to express ${\bf b}^\mu$ in terms of $ {\bf B}$ and ${\bf v}$ only as
\begin{equation}\label{b}  \  {\bf b}^{\mu} = \gamma ( {{\bf B}}/{\gamma ^2}  + {\bf  v} \, ({\bf v} \cdot {\bf B}) \,  ,\, {\bf B}\cdot{ \bf v}) .\end{equation}
Note that in  general   $\partial_\mu {\bf b}^\mu \not =0$ while from Maxwell's equations  we have 
\begin{equation}\label{div}  \partial_\mu  {\bf G }^{\mu\nu} = 0,  \end{equation}
and thus 
\begin{equation}\label{divb}  \partial_\mu  {\bf b}^\mu =  {\bf G }^{\mu\nu}  (\partial_\mu  {\bf u}_\nu )=    {\bf b}^\nu (\partial_\tau {\bf u}_\nu ) ,  \end{equation}
where $ \partial_\tau = {\bf u}^\mu \partial_\mu$\, with $\tau$  the proper time and $\partial_\tau {\bf u}_\nu $  the 4-acceleration of the plasma element.
\\
Unless explicitly stated in the rest of this article we will assume ${\bf e}_\mu \equiv 0$.

{\section{Covariant magnetic 4-vector field equation}\label{frob}

From Eqs.(\ref{2b}) and (\ref{div}) we obtain the ``magnetic vector field equation''  
\begin{equation}\label{meq1}
 {\bf u}^\mu  \partial_\mu {\bf b}^\nu  -    {\bf b}^\mu  \partial_\mu  \,{\bf u}^\nu   + {\bf b}^\nu  \partial_\mu   {\bf u}^\mu  - {\bf u}^\nu  \partial_\mu {\bf b}^\mu  = 0,  \end{equation} 
i.e.
\begin{equation}\label{meq2}
\partial_\tau {\bf b}^\nu = {\bf u}^\nu  \partial_\mu {\bf b}^\mu  \,  - {\bf b}^\nu  \partial_\mu   {\bf u}^\mu  +   {\bf b}^\mu  \partial_\mu  \,{\bf u}^\nu    \end{equation} 
which differs ``in form''  from the standard 3-D magnetic field equation, as obtained e.g. from Eq.(\ref{int3}) by expanding the $ \nabla\times ({\bf v} \times {\bf B}) $  term, because ${\bf b}^\mu$ is not divergence-free.
Inserting  Eq.(\ref{divb})  into Eq.(\ref{meq2})  we obtain the magnetic equation  in relativistic Lagrangian variables 
\begin{equation}\label{meq3}
\partial_\tau {\bf b}^\nu = {\bf u}^\nu   {\bf b}^\alpha (\partial_\tau {\bf u}_\alpha)     \,  - {\bf b}^\nu  \partial_\mu   {\bf u}^\mu  +   {\bf b}^\mu  \partial_\mu  \,{\bf u}^\nu 
 \end{equation} 
where only 3 components are independent.
Eq.(\ref{meq3}) can also be written in projection form as 
\begin{equation}\label{meq4}
(\delta^{\nu}_{\,  \alpha} + {\bf u}^\nu {\bf u}_\alpha) \, \partial_\tau {\bf b}^\alpha =     {\bf b}^\mu  \partial_\mu  \,{\bf u}^\nu - {\bf b}^\nu  \partial_\mu   {\bf u}^\mu  .
 \end{equation} 
 
\subsection{Frobenius condition and 2-D Hypersurfaces}\label{CH}
 
 Equation (\ref{meq1})  can be viewed as
  a Frobenius involution condition for the 4-vector fields   ${\bf b}^\mu$ and ${\bf u}^\mu$.\,
  This condition, which is a consequence of the homogeneous Maxwell's equations $\partial_\mu {\bf G }^{\mu\nu}=0 $, i.e. of Faraday's equation and of ${\bf B}$ being divergence-free, and of   the ideal Ohm's law, allows us\footnote{Provided ${\bf b}^\mu \not =0$, see Sec.\ref{null}} to construct in  the 4-D space-time   2-D hypersurfaces  generated by the vector fields  ${\bf u}^\mu$ and ${\bf b}^\mu$.\\
These hypersurfaces, which we  call {\it connection hypersurfaces} because they will allow us to recast the connection theorem (\ref{int4}) in a covariant form, see Sec.\ref{CovCon}, are  the  4-D counterpart of magnetic field lines in 3-D space and are not related  to the magnetic surfaces  defined in 3-D by the equation ${\bf B}\cdot \nabla \psi = 0$.

\section{Gauge freedom  in the LA representation}\label{GFR}

We can assume, without loss of generality,  that the velocity 4-vector ${\bf u}^\mu$ satisfies a continuity equation of the form
\begin{equation}\label{cont} \partial_\mu(n {\bf u}^\mu) =0,   \end{equation} 
where $n $ can be taken to play the role of the proper density of the plasma element and $n {\bf u}^\mu$ of the  density 4-vector.\\
As shown in detail in \cite{D'A} a gauge freedom is allowed in the definition of the magnetic 4-vector  field ${\bf b}^\mu$ in the LA representation  provided  we relax the orthogonality condition ${\bf b}^\mu {\bf u}_\mu = 0$:
\begin{equation}\label{gauge }    {\bf b}^\mu \rightarrow  {\bf h}^\mu \equiv {\bf b}^\mu +  g \, {\bf u}^\mu ,  \end{equation}
where  $g$ is a  free scalar field.  
\,
Different choices of  the gauge field $g$ allow us to impose  specific conditions on ${\bf h}^\mu$.
If, as in \cite{D'A}, we choose  the {\it divergence gauge} 
\begin{equation}\label{gauge1}   \partial_\tau(g/n)  = - (1/n) \, \partial_\mu {\bf b}^\mu ,  \end{equation}
we  have $\partial_\mu  {\bf h}^\mu = 0$, while, if we take in a given frame\footnote{The quantity $ -  {\bf v}\cdot{ \bf B} $ is a Lorentz scalar. Its expression in a frame moving with respect to the chosen frame with velocity 4-vector $V_\mu$ is $-  (V_\mu b^\mu)/(V_\nu u^\nu)$. } the {\it magnetic gauge} 
\begin{equation}\label{gauge2}  g   = -  {\bf v}\cdot{ \bf B} , \end{equation}
we can make the time component of ${\bf h}^\mu$ vanish and   ${\bf h} \, || \, { \bf  B}$ in that frame.
\\ Note that, since the expression for ${\bf G }^{\mu\nu}$ is unchanged if we insert ${\bf h}^\mu$ for ${\bf b}^\mu$ in Eq.(\ref{2b}),  the Frobenius condition (\ref{meq1})  holds independently of the gauge.
Thus  the connection-hypersurfaces  generated by the 4-vector fields  ${\bf u}^\mu$ and ${\bf b}^\mu$ can also be seen as generated by the 4-vector fields ${\bf u}^\mu$  and ${\bf h}^\mu$. \, 
For the sake of notational clarity in the following  we will denote by ${\bf h}_{||}^{\mu}$  the 4-vector  field corresponding to the  magnetic gauge  (\ref{gauge2}) and  specifically  by ${\bf h}^\mu$  without any additional mark   the 4-vector field  corresponding to the  divergence gauge   (\ref{gauge1}).

\section{Covariant Connection theorem} \label{CovCon}

Extending  the procedure developed in Sec.\ref{ROL}  to  4-D Minkowski space, we consider  in a given frame  a   magnetic field line $\ell $ at a fixed time in 4-D Minkowski space   with  tangent (spacelike) 4-vector  field $d {\bf l}^\mu$. 
\,  In this  frame  its time component $d { l}^o = 0 $   and the  condition {${\bf F}_{\mu\nu}  d {\bf l}^\nu  = 0 $  \,\,  implies $d {\bf  l} \times {\bf B} =0$ corresponding \footnote{It includes
 $d {\bf  l} \cdot {\bf E} =0$ which is satisfied  if  the ideal Ohm law holds}  to the field line condition used in Sec.{\ref{ROL}. \\ Recalling that the rank of 
${\bf F}_{\mu\nu}$ must be even, the condition ${\bf F}_{\mu\nu}  d {\bf l}^\nu  = 0 $  also implies that  $d {\bf l}^\mu $ must be a linear combination of ${\bf b}^\mu$ and ${\bf u}^\mu$ (aside for the null points of ${\bf F}_{\mu\nu}$, see Sec.\ref{null}) i.e.  that it lies on a
connection hypersurface  defined in Sec.\ref{CH}.

\subsection{Time resetting gauge} \label{TG}

The   condition $ {\bf F}_{\mu\nu}  d {\bf l}^\nu= 0 $    remains valid  even without imposing $ d {l}^o =0$ because of the ``time  gauge'' freedom 
 {$ d {\bf l}^\mu  \to d{\hat {\bf l}}^\mu = d{\bf l}^\mu \,  + \,  {\bf u}^\mu \,  d \lambda $},  with  $\lambda$ a scalar function, i.e. $d {\hat {\bf l}}^\mu$ remains in the hypersurface
 generated by ${\bf b}^\mu$ and ${\bf u}^\mu$. \\ Conversely, in a boosted frame (where quantities are denoted by a ``prime'') the transformed vector field  $d {\bf l}^{\prime \mu}$ will acquire a time component but  will still lie on the boosted 2-D hypersurface generated by the boosted vector fields
${\bf b}^{\prime \mu}$ and ${\bf u}^{\prime \mu}$. Then, using the time gauge in reverse as done in \cite{FPEPL},  it will be possible to set $ d { l}^{\prime o} =0$  without violating the condition in the boosted frame  $ {\bf F}^{\prime}_{\mu\nu}  d {\bf l}^{\prime \nu}= 0 $  because of the ideal Ohm's law.

\subsection{Magnetic gauge} \label{MG}
After performing the time resetting gauge, using   the { magnetic gauge}  given by Eq.(\ref{gauge2})  we  can bring the boosted 4-vector field ${\bf b}^{\prime \mu}$  to the form $ {\bf h}^{\prime \mu}_{||} =   (0,  {\bf B}^{\prime}/\gamma ) $.\,
Then  in the boosted frame {$  {\bf F}^{\prime}_{\mu\nu}  d {\bf l}^{\prime \nu}= 0 $  implies  ${\bf d l}^{\prime} \times {\bf B}^{\prime} = 0$.}} 

{\it This proves  that it is possible to define magnetic connections in a covariant way, provided 
we refer to connection hypersurfaces  instead of connection field lines  and  provided we properly  ``gauge''  the 4-vector magnetic field ${\bf b}^\mu$  and the tangent (spacelike) 4-vector  field $d {\bf l}^\mu$  within  the connection hypersurface 
in order to compensate  for the mixing between 3-D magnetic and electric fields under a Lorentz boost  and  for the loss of simultaneity  in different frames.}\\
 Magnetic connections in 3-D space can then be recovered in any chosen reference frame by taking sections of these surfaces at a fixed (in that frame) time.

\section{Coordinates on Connection Hypersurfaces} \label{Com}

Choosing instead  the divergence gauge  (\ref{gauge1}), the 
Frobenius condition (\ref{meq1}) can be reformulated in a way that allows us to define the following two commuting operators
 \begin{equation}\label{0com}   \partial_\tau =  ({\bf n}^\mu/n) \partial_\mu \quad {\rm and} \quad 
\partial_h=  ({\bf h}^\mu/n) \partial_\mu , \end{equation} 
where 
 \begin{equation}\label{1com} \partial_\tau  \partial_h -  \partial_h \partial_\tau    =  [({\bf n}^\mu/n) \partial_\mu   ({\bf h}^\nu/n) \   -   ({\bf h}^\mu/n) \partial_\mu   ({\bf n}^\nu/n)] \, \partial_\nu   \end{equation}
$$= (1/n)  [\partial_\mu   ({\bf n}^\mu {\bf h}^\nu/n   -   {\bf h}^\mu {\bf n}^\nu/n) ]  \, \partial_\nu  \, =  \, (1/n)  [\partial_\mu  {\bf G }^{\mu\nu}] \, \partial_\nu  =  0 . $$
Then the set of curves on a Connection Hypersurface  with tangent fields ${\bf n}^\mu/n = {\bf u}^\mu $ and ${\bf h}^\mu/n$  define a (nonorthogonal, and  in general  only local) coordinate system on the connection hypersurface. From the  Minkowski line element  $ds^2 = d{\bf x}^\mu \eta _{\mu\nu}  d{\bf x}^\nu $ we obtain the following expression for line element on a connection hypersurface 
 \begin{equation}\label{2com}  ds^2 = -d\tau\,^2 + ({\bf h}_\mu {\bf h}^\mu /n^2) \, dh\, ^2   - (2g/n) \,  d h \, d\tau, \end{equation}
where  ${\bf h}_\mu {\bf h}^\mu =  {\bf b}_\mu {\bf b}^\mu  - g^2$ and the gauge function $g$ is defined by Eq.(\ref{gauge1}).

 \section{Advected magnetic 4-D Nulls} \label{null}

When using the LA representation to construct the  connection-hypersurfaces we have not considered the null points  of the e.m. tensor ${\bf F}_{\mu\nu}$ explicitly. 
Note that at these 4-D null points both the magnetic and the electric field vanish, which is a frame independent condition.\\
Since the velocity 4-vector ${\bf u}^\mu$  has no nulls (${\bf u}^\mu {\bf u}_\mu = -1$), in the LA representation with ${\bf e}_\mu = 0$ a 4-D  null of ${\bf F}_{\mu\nu}$ implies a 4-D null of ${\bf b}^\mu$ and {\it viceversa} a null of ${\bf b}^\mu$ implies\footnote{Such a one to one relationship is not generally true  in the case where ${\bf e}_\mu\not=0$ where,  e.g., a null of ${\bf B}$  at the $X$-point of a reconnecting magnetic field  does not imply a null of ${\bf b}_\mu $.}  a null of ${\bf B}$.\\
A generic local expansion around a null 4-point (placed at the origin of the coordinate system)
truncated at the first term reads
\begin{equation}\label{N1} {\bf b}^\mu = N^{\mu}_{\,\, \nu} \,  {\bf x}^\nu ,\qquad   {\rm where  } \quad {\bf u}_\mu {\bf b}^\mu  =0 \, ~ \, \Rightarrow  \, ~ \,  {\bf u}_\mu  N^{\mu}_{\,\, \nu} = 0 .\end{equation}
Here $  N^{\mu}_{\,\, \nu}$ is a numerical tensor  that in general  need not  be symmetric\footnote{Its antisymmetric part is related to the current density 4-vector at the null point}. Then 
\begin{equation}\label{N2}  {\bf G }^{\mu\nu}  = ( N^{\nu}_{\,\, \alpha}   {\bf u}^\mu -  N^{\mu}_{\,\, \alpha} {\bf u}^\nu)\, {\bf x}^\alpha \quad {\rm and } \quad 
\partial_\mu {\bf G }^{\mu\nu} = 0 \,  \Rightarrow \,  N^{\nu}_{\,\, \mu}   {\bf u}^\mu -  N^{\mu}_{\,\, \mu} {\bf u}^\nu =  0 \end{equation}   at the null.\,  Contracting the latter identity with ${\bf u}_\nu$ and using ${\bf u}_\nu N^{\nu}_{\,\,\mu} = 0$ from the r.h.s. of Eq.(\ref{N1}),
we find   at the 4-D null ${N}^{\mu}_{\,\, \mu} = 0 $,  i.e.\,   $\partial _\mu {\bf b}^\mu = 0$ at the null   and \,   $\, N^{\nu}_{\,\, \mu} {\bf u}^\mu  =0. $ \,
Thus in the instantaneous  local rest frame  of the 4-D null  ${N}^{\mu}_{\,\, \nu}$ reduces to  a 3-D tensor (only  its space-space components do not vanish). \\
Finally we note that, since $\partial _\mu {\bf b}^\mu = 0$ at the null point, from Eq.(\ref{meq2}) we find that at the null point $\partial_\tau {\bf b}^\mu =0$, which can be used to trace along the fluid element trajectory the singularities  of the connection-hypersurfaces that arise  at the nulls of ${\bf b}^\mu$.
\\Furthermore, because of the two above conditions at the null point, we can take the gauge function $g$ in Eq.(\ref{gauge1}) equal to zero at the null point so that a null of ${\bf b}^\mu$ corresponds to a null of ${\bf h}^\mu$.  Conversely, since  ${\bf b}^\mu$ cannot be equal to $g {\bf u}^\mu$ with $g\not=0$ because of the orthogonality condition ${\bf u}_\mu {\bf b}^\mu  = 0$, a null point of ${\bf h}^\mu$  must correspond to a null point of ${\bf b}^\mu$ and of $g$.

\section{Covariant Magnetic Helicity}\label{Hel}

As is well known, the homogeneous Maxwell equation  $\partial_\mu {\bf G }^{\mu\nu}  =0 $  implies that we can introduce a 4-vector potential field ${\bf A}_\mu$ such that ${\bf F}_{\mu\nu} = \partial_\mu {\bf A}_\nu -   \partial_\nu {\bf A}_\mu $. 
The introduction of the vector potential  allows us for a general e.m. field   to give 
a  covariant definition of the 4-vector magnetic helicity in the form 
\begin{equation}\label{mh1}
{\bf K}^\mu \equiv {\bf G }^{\mu\nu} {\bf A}_\nu ,   \end{equation} 
such  that \begin{equation}\label{mh2}
\partial_\mu  {\bf K}^\mu  = - {\bf F}_{\mu\nu} {\bf G }^{\nu\mu}/2 .   \end{equation} 
The magnetic helicity 4-vector is defined up to a 4-divergence 
$\partial_\nu ( \chi \, {\bf G }^{\mu\nu})$, with $\chi$ a scalar field, because of the usual gauge freedom in the definition of the 4-vector potential ${\bf A}_\nu \rightarrow {\bf A}_\nu + \partial_\nu \chi$.
\\The r.h.s.  of Eq.(\ref{mh2}) vanishes if the ideal Ohm's law holds.
In this case from Eqs.(\ref{2b})  we find
\begin{equation}\label{mh3}
{\bf K}^\mu  =    {\bf u}^\mu ({\bf b}^\nu   {\bf A}_\nu) - {\bf b}^\mu ({\bf u}^\nu {\bf A}_\nu) ,
  \end{equation} 
i.e.  ${\bf K}^\mu  $ lies on   connection-hypersurfaces.
From the conditions ${\bf F}_{\mu\nu} {\bf u}^\nu =0$ and  ${\bf F}_{\mu\nu} {\bf b}^\nu  =0$, 
%and from the definition of $ {\bf F}_{\mu\nu}$ in terms of the vector potential ${\bf A}_\mu$, 
we obtain  
\begin{equation}\label{mh4}
{\bf u}^\mu\partial_\mu {\bf A}_\nu = {\bf u}^\mu \partial_\nu {\bf A}_\mu, \qquad    {\bf b}^\mu\partial_\mu {\bf A}_\nu = {\bf b}^\mu \partial_\nu {\bf A}_\mu ,   \end{equation} 
and,  using  $\partial_\mu {\bf G }^{\mu\nu}=0$ and Eq.(\ref{mh4}),  we  verify that 
\begin{equation}\label{mh5}
\partial_\mu  {\bf K}^\mu  =  {\bf G }^{\mu\nu} \partial_\mu {\bf A}_\nu  = 
({\bf u}^\mu {\bf b}^\nu - {\bf b}^\mu {\bf u}^\nu) (\partial_\mu {\bf A}_\nu) = {\bf b}^\mu {\bf u}^\nu \partial_\mu {\bf A}_\nu -  {\bf b}^\nu {\bf u}^\mu \partial_\nu {\bf A}_\mu\equiv 0.
 \end{equation} 
If we choose the gauge scalar function $\chi$ such that $ {\bf A}_\mu {\bf u}^\mu = 0$,  i.e. if we make  the time component of the 4-vector potential vanish in the local rest frame, 
from Eqs.(\ref{mh3}), (\ref{mh5})  and  (\ref{cont})  we obtain 
\begin{equation}\label{mh6} \partial_\mu  [{\bf n}^\mu ({\bf b}^\nu   {\bf A}_\nu)/n]   =   {\bf n}^\mu   \partial_\mu ({\bf b}^\nu   {\bf A}_\nu/n) =  0 \quad \rightarrow \quad \partial_\tau ({\bf b}^\nu   {\bf A}_\nu/n) =  0 ,
 \end{equation} 
which provides us with a Lagrangian invariant scalar field advected by the plasma flow.

\section{Conclusions}\label{conc}

In this article we have addressed the problem of defining  covariant magnetic connections  for a relativistic plasma that obeys the ideal Ohm's law and have obtained the following  main results.
\medskip

1) We have  reformulated
the covariant connection theorem discussed in \cite{FPEPL}  in terms of 2-D hypersurfaces in 4-D  Minkowski  space making use  of \\
a)  the representation of the electromagnetic  field tensor in terms of two 4-vector fields  (which we called the Lichnerowicz-Anile representation)  
in the case where the ideal Ohm's law holds,  \\ b)  the  gauge freedom  in the definition  of the magnetic 4-vector  field, \\ c)   a time-gauge transformation (time resetting) of the  4-vector  field tangent to the curve connecting two plasma elements in 4-D space.\\    We call  these hypersurfaces connection  hypersurfaces.

2)  We have indicated that these connection  hypersurfaces  take the role, for the full electromagnetic  field tensor in 4-D,  of  the magnetic field lines in 3-D. \\ We thus argue that  a covariant definition of magnetic reconnection may be given in a 4-D framework as a local  ``piercing and merging'' of connection hypersurfaces that lose  their  identity   only locally (where ${\bf e}_\mu \not = 0$ and the Frobenius condition does not hold), just as  magnetic field  lines do in the standard 3-D space  setting.
\medskip

Regarding point 1), we stress that   different forms of gauge freedom   play  a very important role in our  formulation,  a common feature of electrodynamic theory. In fact the use of  gauge transformations  is a convenient tool for  implementing  useful but non-explicitly covariant conditions in  a covariant theory.  A well known example is provided by the transverse potential gauge condition ($\phi  = 0$, with $\phi$ the time component of the vector potential) for a plane electromagnetic wave. This condition is  not explicitly covariant, i.e. it is not  in general preserved by a Lorentz boost, but can be restored by a gauge transformation of the boosted vector potential.

Point 2) suggests  that the  investigation of  topological  properties  of the magnetic field in 3-D space,  which  play a fundamental role in  ideal MHD,  should be extended to the  investigation of  the topological  properties  of the full electromagnetic field  tensor in 4-D space.
This  future line of enquiry 
may well open a novel and rich way of reinterpreting the topological properties of ideal MHD.

\medskip

Finally we note again that the treatment developed in the present  paper   does not involve  the full  set  of MHD plasma equations  and only requires that an ideal Ohm's law in terms of  a  fluid velocity be satisfied. 
Thus  this treatment can be  applied to different  plasma theories where  the velocity 4-vector  field ${\bf u}^\mu$ is not the plasma fluid velocity but, for example,  the electron fluid velocity as is the case, e.g.,  in EMHD (see \cite{EMHD}).

\section*{Acknowledgments}
The author acknowledges  clarifying   discussions with G. Tomassini, P.J. Morrison and E. D'Avignon.

% Note the spaces between th

\bibliography{jpp-instructions}

\end{document}